\documentclass{iau}
\usepackage{graphicx,natbib}
 
\newcommand{\farcs}{$.\!\!^{\prime\prime}$}
\newcommand{\kms}{\,km\,s$^{-1}$}
\newcommand{\MO}{M$_\odot$\,yr$^{-1}$}

\title{The Narrow Line Region in 3D: mapping AGN feeding and feedback}

\author[Storchi-Bergmann]{Thaisa Storchi Bergmann$^{1, 2} $}

\affiliation{$^1$Instituto de F\'isica,UFRGS, Campus do Vale, Porto Alegre, RS, Brazil\\
 $^2$Visitor Professor at the Harvard-Smithsonian Center for Astrophysics\\
email: {\tt thaisa@ufrgs.br} \\
}

\pubyear{2014}
\volume{309}
\jname{Galaxies in 3D across the Universe}
\editors{B. L. Ziegler, F. Combes, H. Dannerbauer, M. Verdugo, eds.}

\begin{document}

\maketitle

\begin{abstract}
Early studies of nearby Seyfert galaxies have led to the picture that the Narrow Line Region is a cone-shaped region of gas ionized by radiation from a nuclear source collimated by a dusty torus, where the gas is in outflow. In this contribution, I discuss a 3D view of the NLR obtained via Integral Field Spectroscopy, showing that: (1) although the region of highest emission is elongated (and in some cases cone-shaped), there is also lower level emission beyond the ``ionization cone", indicating that the AGN radiation leaks through the torus; (2) besides outflows, the gas kinematics include also rotation in the galaxy plane and inflows; (3) in many cases the outflows are compact and restricted to the inner few 100pc; we argue that these may be early stages of an outflow that will evolve to  an open-ended, cone-like one. Inflows are observed in ionized gas in LINERs, and in warm molecular gas in more luminous AGN, being usually found on hundred of pc scales. Mass outflow rates in ionized gas are of the order of a few \MO, while the mass inflow rates are of the order of tenths of \MO. Mass inflow rates in warm molecular gas are $\approx$\,4--5 orders of magnitude lower, but these inflows seem to be only tracers of more massive inflows in cold molecular gas that should be observable at mm wavelengths.

\end{abstract}

\firstsection
\section{Introduction}

The physical processes that couple the growth of supermassive black holes (SMBH) to their host galaxies 
occur in the inner few kpc \citep{hopkins10} when the nucleus becomes active due to mass accretion to the SMBH \citep{ferrarese05,KH13}. The radiation emitted by Active Galactic Nuclei (AGN) ionizes the gas in the vicinity of the nucleus, forming the Narrow Line Region (NLR). AGN-driven winds \citep{ciotti10,elvis00} interact with the gas and produce outflows that are observed in the NLR reaching velocities of hundreds of km\,s$^{-1}$ \citep{sb10,fisher13}. Relativistic jets emanating from the AGN also interact with the gas of the NLR \citep{wang09}. Both types of outflow produce feedback, which is a necessary  ingredient in galaxy evolution models to avoid producing over-massive galaxies \citep{fabian12}. The importance of the NLR stems from the fact that it is spatially resolved, extending from hundreds to thousands of parsecs from the nucleus and exhibits strong line emission. These properties  allow the observation  of the interaction between the AGN and the circumnuclear gas in the galaxy:  (1) via the observation of the NLR geometry and excitation properties that constrain the AGN structure and ionizing source; (2) via the gas kinematics that maps the processes of the AGN feeding and feedback.

In this contribution, I show results from a ``3D" study of the NLR of nearby AGNs by our research group AGNIFS (AGN Integral Field Spectroscopy). I present maps of the gas flux distribution and kinematics, as well as estimates of gas mass inflow and outflow rates, from observations of both the ionized and hot molecular gas.\\


We have used two Gemini instruments in these studies: the Near-IR Integral Field Spectrograph NIFS \citep{mcgregor03}, with the adaptative optics module ALTAIR and the Gemini Multi-Object Spectrograph Integral Field Unit (GMOS-IFU). With NIFS, we have performed observations in the Z, J, H and K bands of the near-IR, at an angular resolution of about 0\farcs1 and spectral resolution R$\approx$5300. With the  GMOS-IFU, we have mostly observed the spectral region covering the emission lines [O\,I]$\lambda$6300, H$\alpha$+[N\,II] and [S\,II]$\lambda\lambda$6717,31, at a spectral resolution  R$\approx$2000. The NIFS field-of-view (hereafter FOV) is 3$^{\prime\prime}\times3^{\prime\prime}$, while the GMOS-IFU FOV can be either 3\farcs5$\times5^{\prime\prime}$ in one-slit mode, or 7$^{\prime\prime}\times5^{\prime\prime}$, in two-slit mode. In a few cases we have performed mosaiced observations to cover a larger FOV.

\section{An hourglass-shaped/conical outflow}

\begin{figure}
\centering
\includegraphics[scale=0.26]{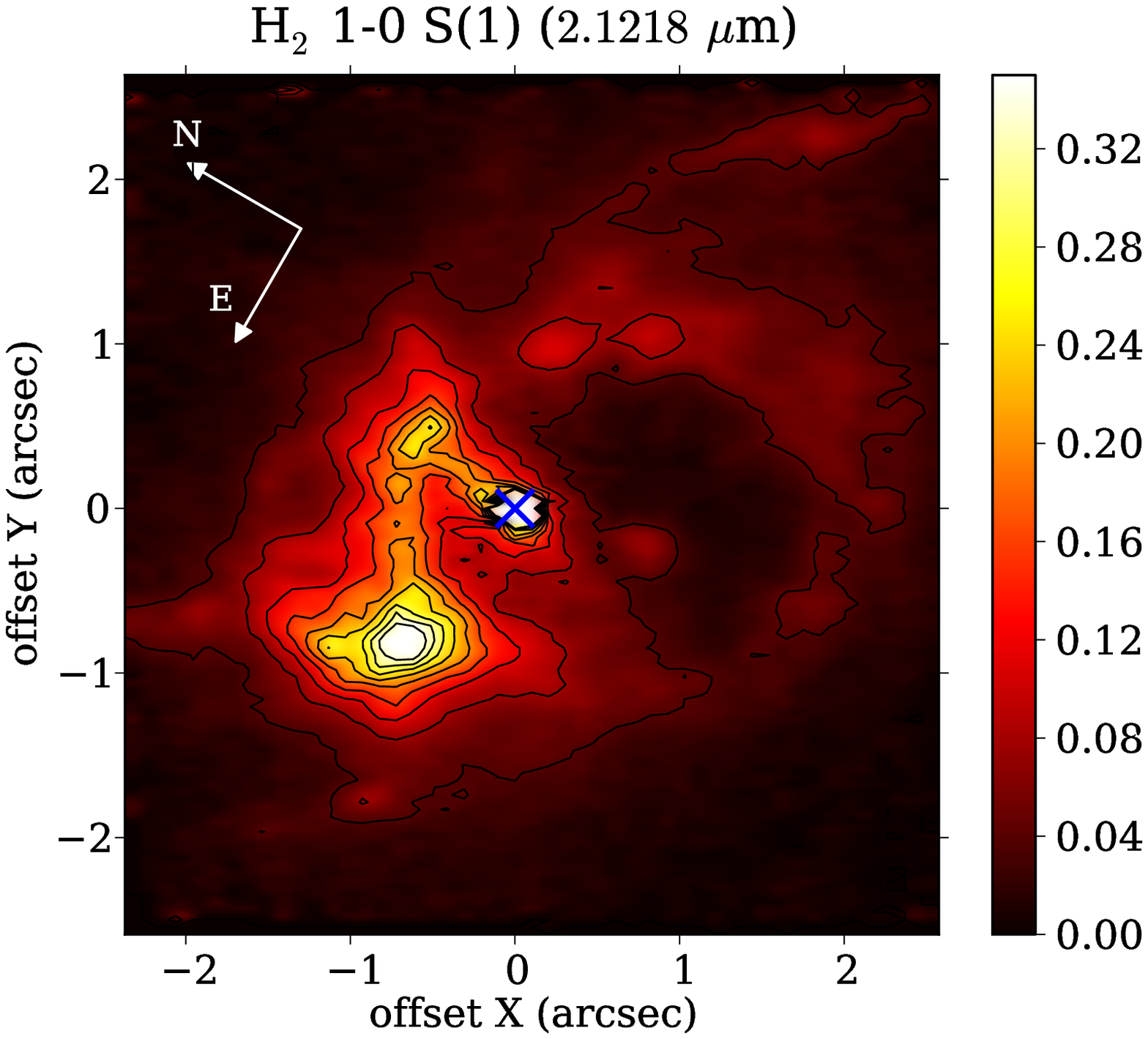} 
\includegraphics[scale=0.26]{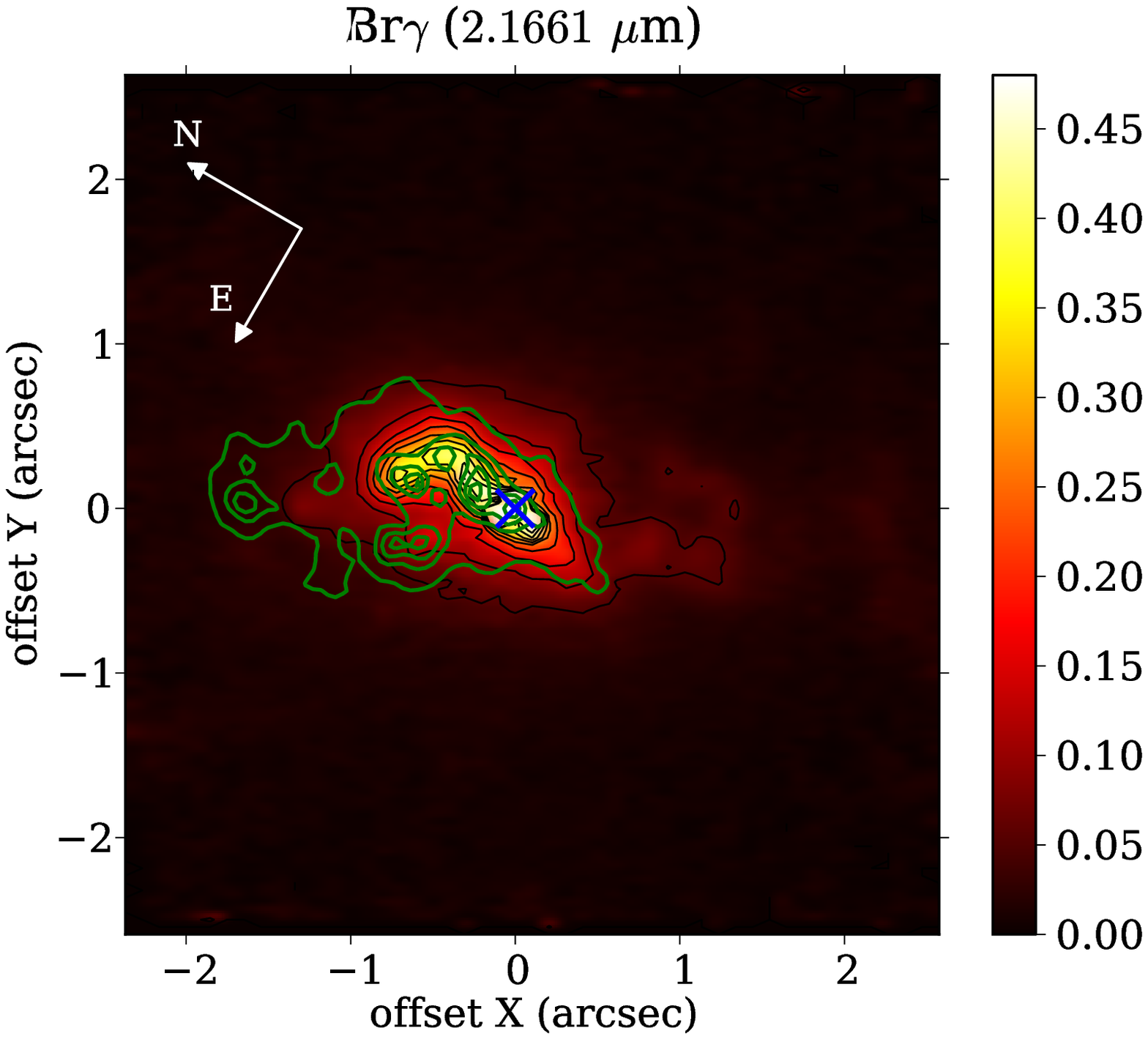}
\includegraphics[scale=0.26]{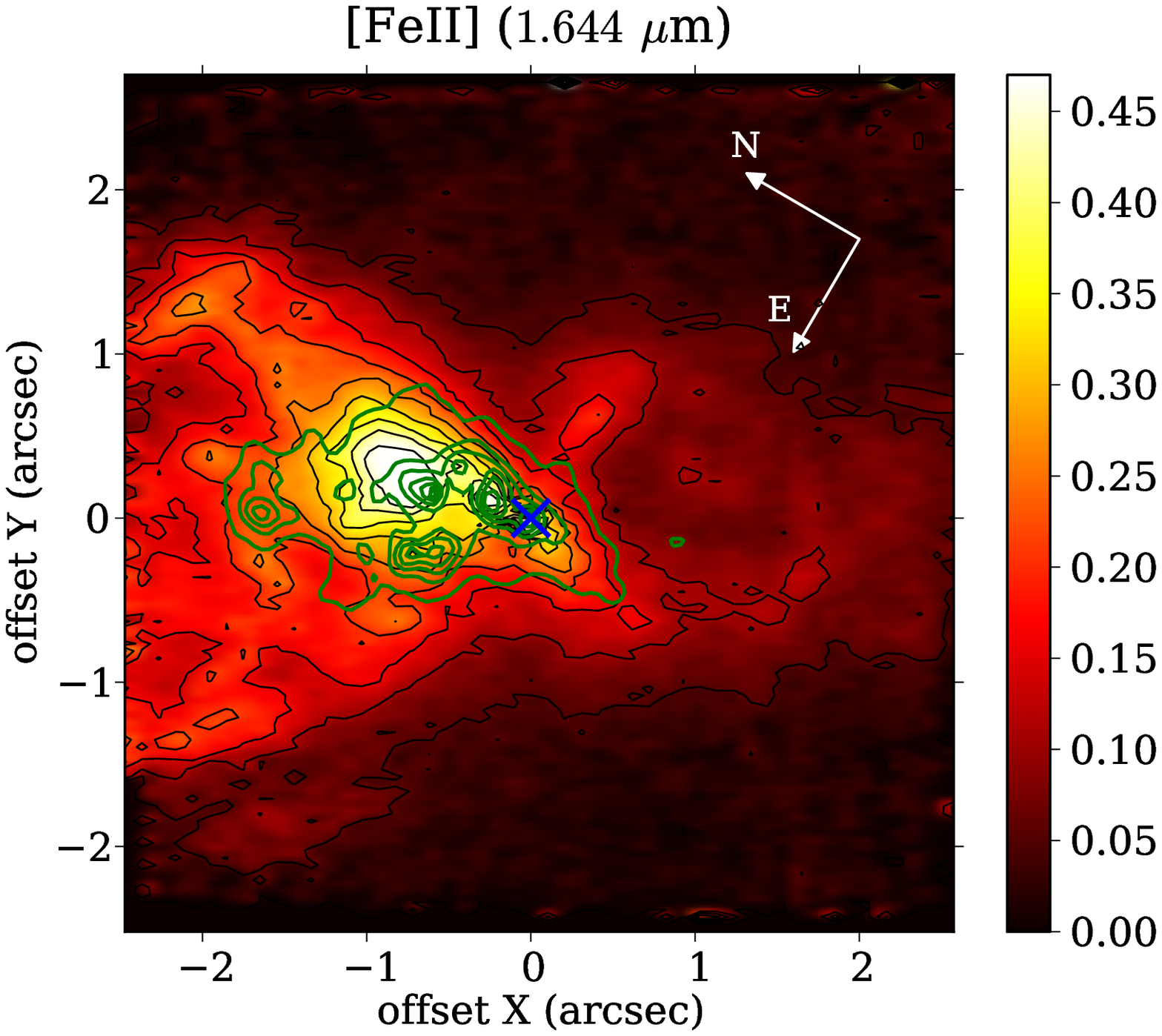} 
\caption{Gas flux distributions in the emission lines H$_2\lambda2.12\mu$m, Br$\gamma$ and [Fe\,II]1.644$\mu$m within the inner $\approx$200\,pc radius of NGC\,1068 from \citet{riffel14a}. The green contours are from the flux distribution in the optical [O\,III]$\lambda$5007 emission line.}\label{n1068_flux}
\end{figure}

I begin with the NLR of the prototypical Seyfert 2 galaxy NGC\,1068, that we have observed with NIFS+ALTAIR in the near-IR. At the adopted distance of 14.4\,Mpc \citep{riffel14a}, the spatial resolution of our observations is $\approx$7\,pc. We illustrate in Fig.\ref{n1068_flux} the flux distributions in the emission lines of H$_2\lambda2.12\mu$m of the warm (T$\approx$2000K, \citet{riffel14a}) molecular gas, compared with those of Br$\gamma$ and [Fe\,II]$\lambda1.644\mu$m. While the warm molecular gas is distributed in a $\approx$100\,pc ring, the Br$\gamma$ flux distribution is very similar to that in the optical [O\,III]$\lambda$5007 emission line (green contours). The rightmost panel shows that the ionized flux distribution in [Fe\,II] is distinct from the two previous ones, showing an hourglass-shaped structure, resembling that seen in some planetary nebulae. We attribute these differences to: (1) the fact that the molecular gas is in the galaxy disc, while both the gas emitting  Br$\gamma$ and [Fe\,II] extends to high galactic latitutes, where they are in outflow; (2) the Br$\gamma$ and [O\,III] emission are restricted to the region of highest ionization, where the collimated radiation from the AGN incides; (3) the [Fe\,II] emission comes from the larger partially ionized region, that extends beyond the fully ionized region, and is thus a better tracer of the whole outflow from the AGN.

In \citet{barbosa14} we show a comparison between channel maps obtained along the [Fe\,II] and H$_2$ emission line profiles. While the [Fe\,II] emission is observed in outflow (following the hourglass shape that extends to high galactic latitudes) up to blueshifts and redshifts of  $\approx$\,800\,\kms, the H$_2$ emission is only observed at low velocities ($\approx$100\kms), consistent with a kinematics dominated by rotation (and expansion in this case, as pointed out in the paper) in the galaxy disk.

The NLR of  NGC\,1068 thus has the expected ``conical" shape, where the gas is in outflow. But these characteristics seem to be more the exception than the rule. In fact, \citet{fisher13}, in a long-slit study of the NLR of a sample of $\approx$60 active galaxies, using the instrument STIS aboard the Hubble Space Telescope, have concluded that in only about 1/3 of their sample, the kinematics of the NLR is consistent with a conical outflow. Although our 3D studies have so far covered a smaller sample, our results support also varied morphologies and velocity fields for the observed outflows.

\section{Other types of outflow}

\begin{figure}
\centering
\begin{minipage}{1\linewidth}
\includegraphics[scale=0.35]{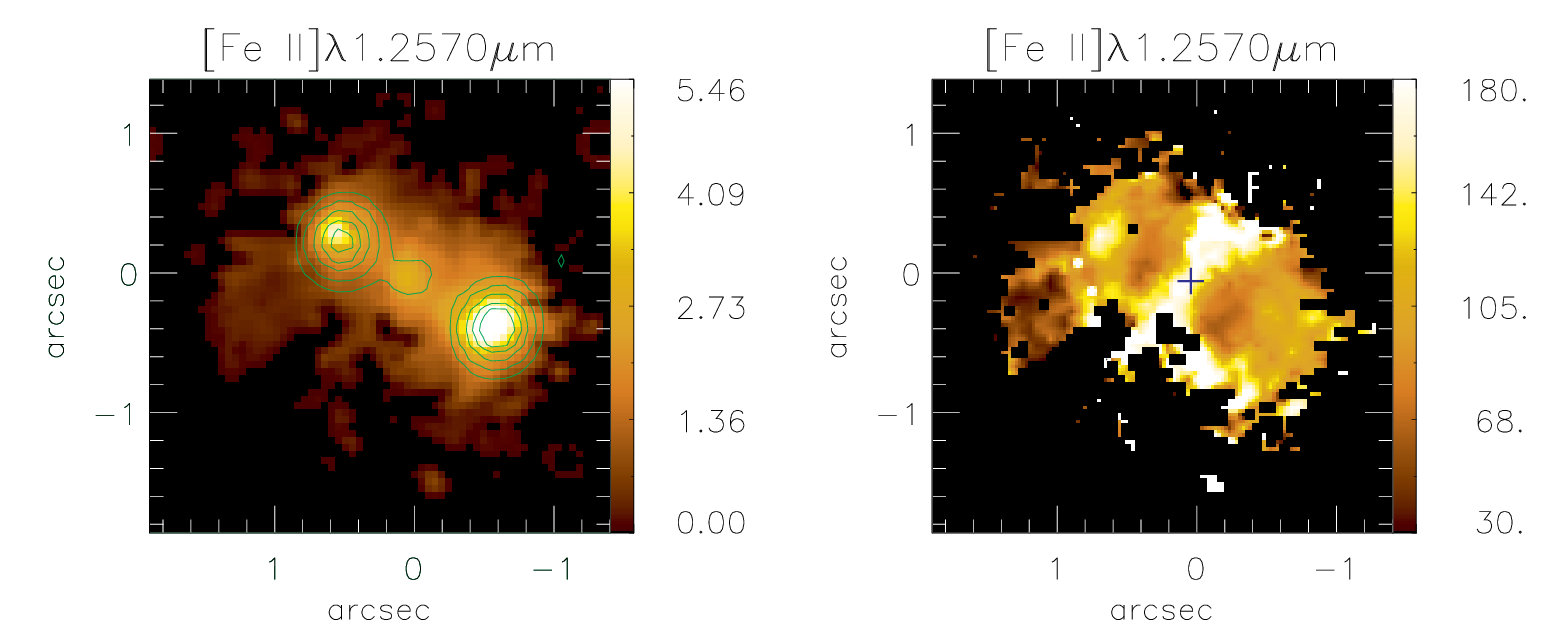} 
\end{minipage}
\begin{minipage}{1\linewidth}
\vspace{-4cm}
\hspace{8cm}
\includegraphics[scale=0.3]{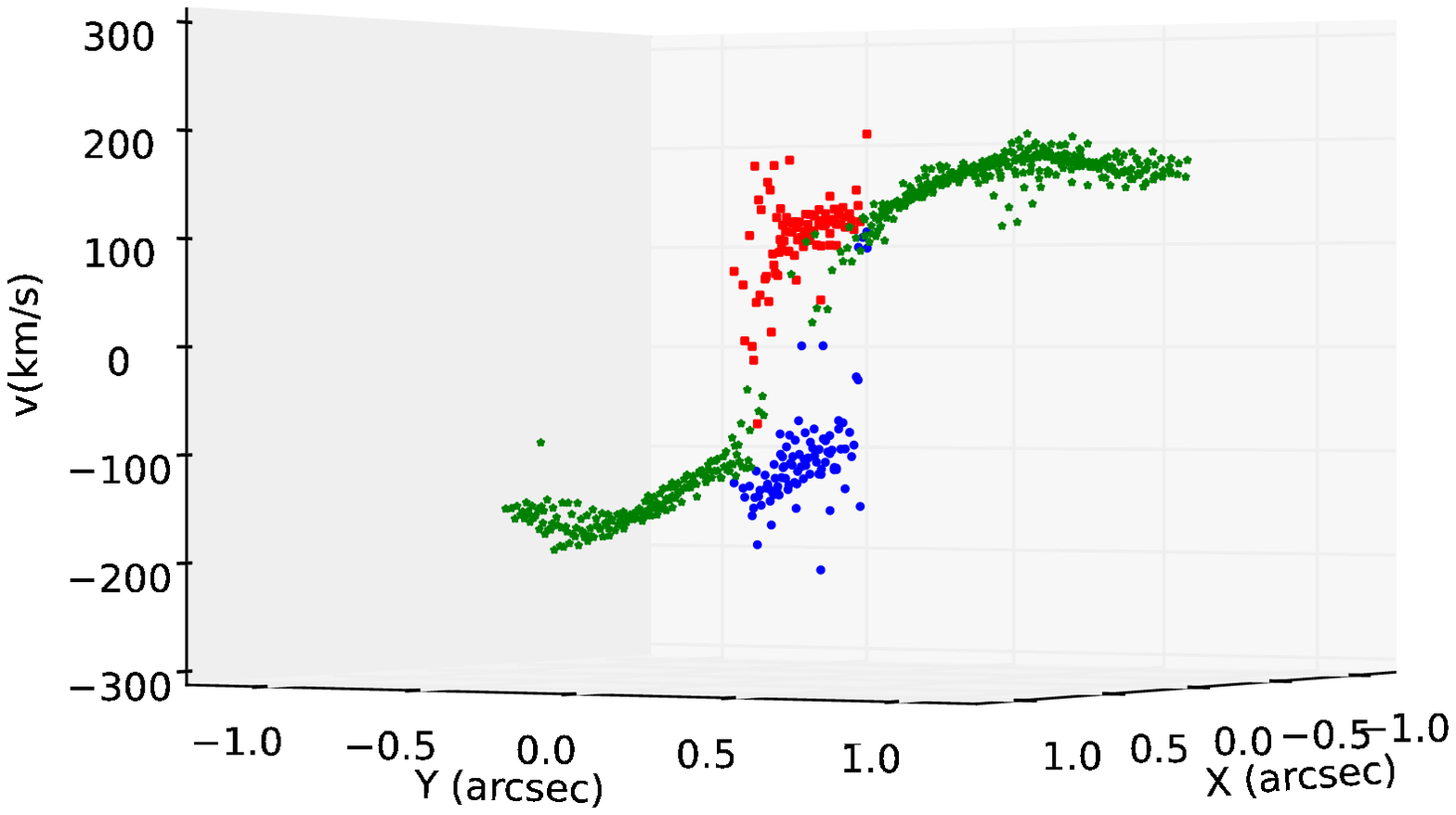}
\end{minipage}
\caption{Left panel: gas flux distribution in the line [Fe\,II]1.644$\mu$m of the inner $\approx$300\,pc of NGC\,5929 \citep{riffel14b}. The green contours are from a radio image. Central panel: [Fe\,II] velocity dispersion, showing a ``stripe" of high velocity dispersion. Right panel: a rendering of the velocity field showing that the highest velocity dispersions correspond to two components (one in blueshift and the other in redshift), interpreted as due to an equatorial outflow.}  \label{n5929}
\end{figure}

Let's consider the case of NGC\,5929 \citep{riffel14b}. Our NIFS observations of the NLR of this Seyfert 2 galaxy allow us to resolve 0\farcs1$=$18\,pc. The left panel of Fig.\ref{n5929} shows the flux distribution in the [Fe\,II]$\lambda1.64\mu$m emission line over the inner $\approx$\,300\,pc, that is most extended along the direction connecting two radio hot spots shown in green in the figure. The central panel shows  a ``stripe" of high velocity dispersion values running perpendicularly to the radio jet. Enhanced velocity dispersion is observed also at the loci of the radio hot spots, where it can be attributed to interaction between the radio jet and the ambient gas.

 A closer look at the emission-line profiles along the high velocity dispersion stripe of Fig.\,\ref{n5929} shows that they are better fitted by two velocity components, one blueshifted and the other redshifted, as illustrated in the rightmost panel of the figure. This 3D rendering of the velocity field shows that most of the gas is rotating with a major axis approximately coincident with the orientation of the radio axis, but there is also an outflow along the equatorial plane of the AGN. In \citet{riffel14b} we have interpreted this as originating in an accretion disk wind running almost parallel to the surface of the disk \citep{proga04} and/or in a wind emanating from the AGN torus \citep{elitzur12}.

We have found outflows perpendicular to the ionization axis in three other cases. In  Arp\,102B \citep{couto13}, observed with GMOS-IFU, the H$\alpha$ emission-line kinematics reveal gas in rotation within the inner kiloparsec, but showing also an outflow perpendicular to the ionization axis and the radio jet within the inner $\approx$\,200\,pc.

In the Seyfert 2 galaxy NGC\,2110, also using GMOS-IFU, we \citep{schnorr14a} found elongated line emission over the whole FOV, that extends to $\approx$\,800\,pc from the nucleus, intepreted as due to ionization by a collimated nuclear source. The gas kinematics reveal that most of this gas is in rotation in the galaxy disk, while within the inner $\approx$\,300\,pc there is a high velocity dispersion component that we have attributed to a nuclear outflow. 

\begin{figure}
\centering
\includegraphics[scale=0.4]{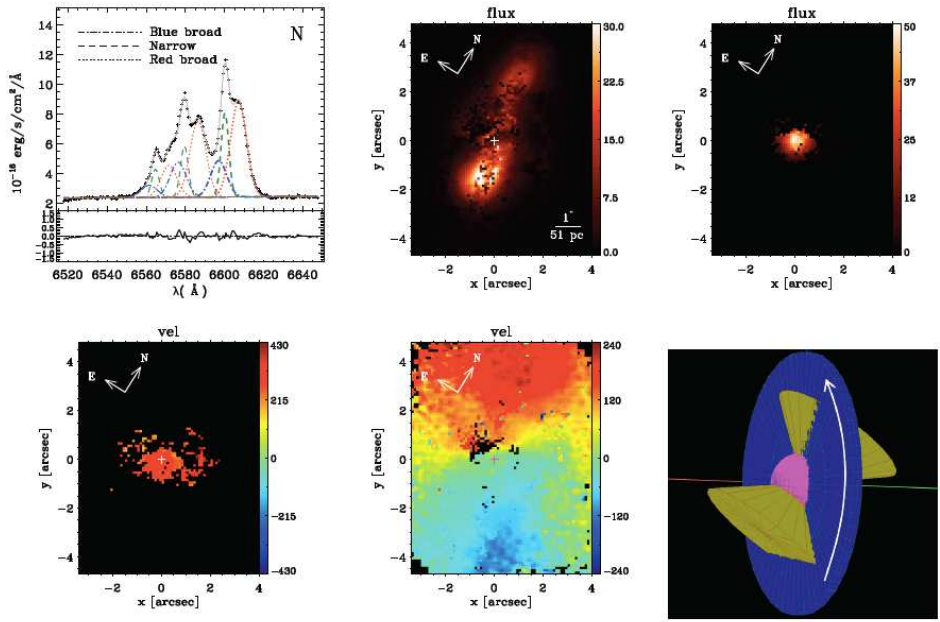}
\caption{Top left: fit of the H$\alpha+$[N\,II] emission line profiles of the inner 50\,pc of NGC\,1386, showing a central narrow plus two broad components (nuclear outflow); top center: [NII]$\lambda$6584 flux distribution of the narrow component, showing the region of strongest emission; top right: flux distribution of the broad components; bottom left: velocity of the red broad component; bottom center: velocity fied of the narrow component; bottom right: scenario of a nuclear outflow (pink) and ionization cones (yellow) that intercept the disk (blue) \citep{lena14}.}  \label{n1386}
\end{figure}

Yet another case of a compact nuclear outflow is observed in the Seyfert 2 galaxy NGC\,1386 \citep{lena14}, where our GMOS-IFU observations covered the inner $\approx$\,500\,pc. We found a similar result to that found in NGC\,2110: while the gas emission is elongated, extending up to the borders of the FOV, and can thus be attributed to ionization by a collimated nuclear source, the kinematics shows again that most of the gas is in rotation in the galaxy disk. Only within the inner $\approx$\,50\,pc there is a region of high velocity dispersion and double components, interpreted as due to a nuclear outflow.  We show this in Fig.\,\ref{n1386}, where the cartoon illustrates our preferred explanation for this case, but the general idea should also apply to the other galaxies with compact outflows discussed above.

\section{Inflows}

The observations of the NLR discussed above have shown that in many cases the gas emission is elongated or cone-shaped but the kinematics is dominated by rotation in the galaxy disk. Only in the innermost region we observe outflows. The AGN thus works as a flashlight that illuminates the ambient gas that is usually rotating in the galaxy disk. This fact has allowed us to see also inflows in the gas, which are usually associated to nuclear spiral arms or dusty filaments seen in structure maps \citep{sl07}. We have observed inflows in ionized gas around LINER nuclei, where outflows are very weak or absent, making it easier to see the inflows. Examples of galaxies where we have observed inflows in ionized gas are: NGC\,1097 \citep{fathi06}, NGC\,6951 \citep{sb07}, M\,81 \citep{schnorr11} and NGC\,7213 \citep{schnorr14b}. 

\begin{figure}
\centering
\includegraphics[scale=0.9]{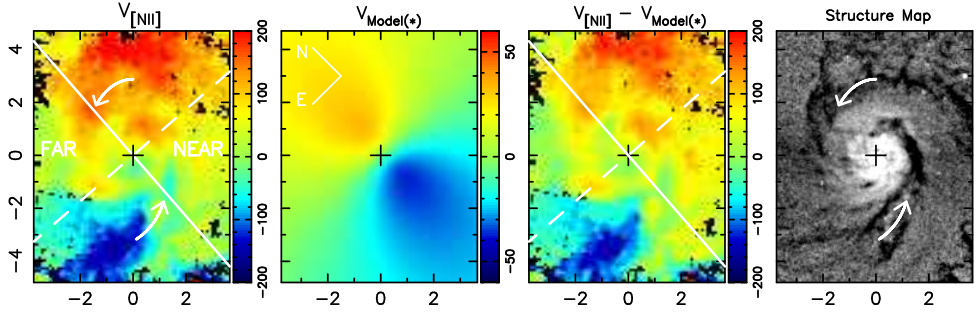}
\caption{Left: gas velocity field of the inner 500\,pc of NGC\,7213; center left: model fitted to the stellar velocity field; center right: residuals between the gas and model stellar velocity field; right: structure map.}\label{n7213}
\end{figure}

The case of NGC\,7213 is illustrated in Fig.\,\ref{n7213}. The nucleus of this galaxy is classified as LINER/Seyfert 1, and we have obtained GMOS-IFU data of the inner $\approx$\,500\,pc. The excellent signal-to-noise ratio of the data has allowed us to derive the stellar kinematics, revealing high velocity dispersion ($\approx$\,200\kms) and rotation
(amplitude of $\approx$50\kms). Gas emission is observed over the whole FOV, presenting also rotation, but with an amplitude much larger than that of the stars. The leftmost panel of Fig.\,\ref{n7213} shows the gas velocity field, that can be compared with the model fitted to the stellar velocity field, shown in the second panel. The third panel (to the right) shows the difference between the velocity field of the gas and that of the stars. Large residuals are found in redshift in the near side of the galaxy and in blueshift in the far side. We note that these residuals follow a spiral pattern that is correlated with a dusty spiral seen in a structure map built from an HST F606W image of the galaxy, shown in the rightmost panel of the figure. Under the assumption that the gas is in the plane of the galaxy (this is justified in the paper), we can only conclude that the excess blueshifts and redshifts are due to gas inflows along the spirals. The estimated mass inflow rate along these spirals are 0.4\,\MO at $\approx$\,400\,pc and 0.1\,\MO at $\approx$\,100\,pc of the nucleus. 

We have found signatures of inflows also in warm molecular gas via 3D observations of the H$_2$ emission line at 2.122$\mu$m in the near-IR using NIFS. This line allows us to observe inflows even in luminous Seyfert galaxies. Our studies indicate that this line is usually excited by X-rays from the AGN, that penetrate in the galaxy disk. The disk origin is supported by the  H$_2$ kinematics that is usually quite distinct from that of the ionized gas also observed in the near-IR. While the ionized gas (e.g. H$^+$ and [Fe\,II]) shows higher velocity dispersion, rotation and outflows, the H$_2$ kinematics shows lower velocity dispersion, rotation and inflows, supporting its location in the galaxy disk. H$_2$ inflows have been found in NGC\,4051 \citep{riffel08}, Mrk\,1066 \citep{riffel11}, Mrk\,79 \citep{riffel13}. In the case of Mrk\,766 \citep{schonell14}  we have found rotation in a compact disk, supporting also that this gas is feeding the AGN. Mass inflow rates in warm molecular gas are 4-5 orders of magnitudes lower than that observed in ionized gas. Nevertheless, the warm molecular gas may only be the heated skin of a much larger cold molecular gas reservoir (and inflow) that should be observable at mm wavelenghts \citep[e.g.][]{combes14}.



\section{Conclusions}

I have reported 3D observations of the NLR of nearby active galaxies using integral field spectroscopy at spatial resolutions of 10--100\,pc. These observations reveal that the AGN radiation works as a flashlight illuminating, heating and ionizing the gas in the circumnuclear regions. Ionized gas is seen not only along the AGN collimation axis, but over all directions (although at a lower intensity), indicating escape of radiation through the walls of the AGN torus. The illuminated, heated or ionized gas may be in outflow, inflow or rotating in the galaxy disk. The outfows are observed following a hollow cone or hourglass geometry or in a compact structure in the inner 50-300\,pc; we suggest that these compact outflows may be the initial phase of an outflow that will evolve to an cone-like outflow. Beyond this region, rotation and inflows in the galaxy disk are observed. In a few LINERs studied, there is no outflow, and only gas rotating and inflowing in the galaxy disk are observed. Outflow velocities are in the range 200 -- 800\kms and the mass outflow rates are of the order of a few \MO.  Inflows are observed along nuclear spirals, usually on $\approx$\,100\,pc scales, at typical mass inflow rates in the range 0.1--1\MO. These inflow rates are much larger than the typical AGN accretion rates ($\approx$\,10$^{-3}$\MO), and will probably lead to the formation of new stars in the circumnuclear region \citep[e.g.][]{sb12}.




\begin{thebibliography}{99}


\bibitem[{{Allington-Smith et al.}(2002)}]{gmos02} Allington-Smith, J. et al. 2002, PASP, 114, 892

\bibitem[{{Barbosa et al.}(2014)}]{barbosa14} Barbosa, F. K. B. et al. 2014, MNRAS, in press

\bibitem[{{Combes et al.}(2014)}]{combes14} Combes, F. et al. 2014, A\&A, 565, 97

\bibitem[{{Couto et al.}(2013)}]{couto13} Couto, G. et al. 2013, MNRAS,435, 2892

\bibitem[{{Ciotti} \& {Ostriker}(2010)}]{ciotti10} Ciotti L, Ostriker JP, Proga D. 2010, ApJ 717, 707 


\bibitem[{{Elitzur}(2012)}]{elitzur12} Elitzur M., 2012, ApJL, 747, L33.

\bibitem[{{Elvis}(2000)}]{elvis00} Elvis, M. 2000, ApJ, 545, 63

\bibitem[{{Fabian}(2012)}]{fabian12}  Fabian, A. C., 2012,  ARAA, 50, 455

\bibitem[{{Fathi et al.}(2006)}]{fathi06} Fathi, K. et al. 2006, ApJ, 641, L25

\bibitem[{{Ferrarese} \& {Ford}(2005)}]{ferrarese05}    Ferrarese, L.  \& Ford, H., 2005, SSRv, 116, 523 

\bibitem[{{Fisher et al.}(2013)}]{fisher13}   Fisher, T.  et al. 2013, ApJS, 209, 1

\bibitem[{{Hopkins} \& {Quataert}(2010)}]{hopkins10} {Hopkins} P. \& {Quataert} E., 2010, MNRAS, 407, 1529 

\bibitem[{{Kormendy} \& {Ho}(2013)}]{KH13}  Kormendy, J. \& Ho, L. C, 2013, ARA\&A, 51, 511

\bibitem[{{Lena et al.}(2014)}]{lena14} Lena, D. et al. 2014, submitted

\bibitem[{{McGregor et al.}(2003)}]{mcgregor03} McGregor, P. J. et al., 2003, Proceedings of the SPIE, 4841, 1581

\bibitem[{{M\"uller S\'anchez et al.}(2009)}]{muller09} M\"uller S\'anchez, F., Davies, R.I., Genzel, R., et al.\ 2009, ApJ,
691, 749

\bibitem[{{Proga} \& {Kallman}(2004)}]{proga04} Proga, D. \& Kallman, T. R. 2004, ApJ, 616, 688

\bibitem[{{Riffel et al.}(2008)}]{riffel08} Riffel, R. et al. 2008, MNRAS, 385, 1129

\bibitem[{{Riffel \& Storchi-Bergmann}(2011)}]{riffel11} Riffel, R. A. \& Storchi-Bergmann, T., 2011, MNRAS, 411, 469.

\bibitem[{{Riffel et al.}(2013)}]{riffel13} Riffel, R. A.; Storchi-Bergmann, T. \& Winge, C. 2013, MNRAS, 430, 2249

\bibitem[{{Riffel et al.}(2014a)}]{riffel14a} Riffel, R. A., Vale, T. B., Storchi-Bergmann, T., McGregor, P. J. 2014, MNRAS, 442, 656

\bibitem[{{Riffel et al.}(2014b)}]{riffel14b} Riffel, R. A., Storchi-Bergmann, T. \& Riffel, R. 2014, ApJ, 780, L24

\bibitem[{{Schnorr M\"uller et al.}(2011)}]{schnorr11} Schnorr-M\"uller, A. et al. 2011, MNRAS, 413, 149

\bibitem[{{Schnorr M\"uller et al.}(2014a)}]{schnorr14a} Schnorr M\"uller, A. et al. 2014, MNRAS, 437, 1708

\bibitem[{{Schnorr M\"uller et al.}(2014b)}]{schnorr14b} Schnorr M\"uller, A. et al. 2014, MNRAS, 438, 3322

\bibitem[{{Schonell et al.}(2014)}]{schonell14} Schonell, A. Jr. et al. 2014, MNRAS, in presss

\bibitem[{{Sim\~oes Lopes et al.}(2007)}]{sl07} Sim\~oes Lopes R. D., Storchi-Bergmann T., Saraiva M. F. \& Martini P., 2007, ApJ, 655, 718.

\bibitem[{{Storchi-Bergmann et al.}(2007)}]{sb07} Storchi-Bergmann, T. et al. 2007, ApJ, 670, 959.

\bibitem[{{Storchi-Bergmann et al.}(2010)}]{sb10}  Storchi-Bergmann, T. et al. 2010, MNRAS, 402, 819

\bibitem[{{Storchi-Bergmann et al.}(2012)}]{sb12}  Storchi-Bergmann, T. et al. 2012, ApJ, 755, 87

\bibitem[{{Wang} \& {Elvis}(2009)}]{wang09}  Wang, J., Elvis, M. et al. 2009, ApJ, 704, 1195



\end{thebibliography}
\end{document}